# Observation of enhanced spin-spin correlations at triple point in 2D ferromagnetic $Cr_2X_2Te_6$ (X=Si, Ge)

Yugang Zhang[1,2†], Zefang Li[3†], Jing Zhang[1,2], Long Zhang[1,2], Yonghong Li[1,2], Shuang Liu[2], Young Sun[2*], Wenhong Wang[3,4*], and Yisheng Chai[1,2*]

[1] *Low Temperature Physics Laboratory, College of Physics, Chongqing University, Chongqing 401331, China*

[2] *Center of Quantum Materials and Devices, Chongqing University, Chongqing 401331, China*

[3] *Beijing National Laboratory for Condensed Matter Physics, Institute of Physics, Chinese Academy of Sciences, Beijing 100190, China.*

[4] *Tiangong University, Tianjin, 300387, China*

[*] youngsun@cqu.edu.cn, wenhongwang@tiangong.edu.cn, yschai@cqu.edu.cn

[†]These authors contributed equally to this work

The domain dynamics and spin-spin correlation of 2D ferromagnets $Cr_2X_2Te_6$ (X=Si, Ge) are investigated by a composite magnetoelectric method. The magnetic field-temperature phase diagrams for both in-plane and out-of-plane magnetic fields disclose a triple point around $T_C$ and 1 kOe, where ferromagnetic, paramagnetic, and spin-polarized phases coexist. The magnetoelectric signal shows peak features at the phase boundaries and reaches the maximum at the triple point, suggesting significant enhancement of spin-spin correlations at this point. A comparison between two systems reveals that the spin-spin correlations in $Cr_2Si_2Te_6$ are stronger than that in $Cr_2Ge_2Te_6$.

Recently, the easy-exfoliated layered 2D magnetic materials such as $CrI_3$ [1], $Cr_2Ge_2Te_6$(CGT) [2], $Cr_2Si_2Te_6$ (CST) [3], and $Fe_3GeTe_2$ [4], have attracted considerable research attention. Even though the Ginzburg criterion states that 2D magnetic systems are much more vulnerable to magnetic fluctuations, low dimensional long-range magnetic order in few atomic layers has already been observed experimentally due to the strong magnetic anisotropy [2]. These materials provide the ideal platform for exploring the strong magnetic fluctuation down to the 2D limit [2,5], and show great promise for future spintronic applications.

Chromium tellurides CST and CGT are uniaxial ferromagnetic (FM) semiconductors [6] and belong to the class of layered transition metal trichalcogenides, which are crystallized in the trigonal space group $\overline{R}3$ (148). Figure 1(a) shows the ABC-stacking CST and CGT layers separated by cleavable van der Waals gaps. $Cr^{3+}$ is responsible for the spontaneous magnetic order at $T_c$ = 34 K and 67 K for CST and CGT, respectively, with the easy axis aligning along the *c* axis [7]. According to the Landau's theory of second order phase transition, there is a critical point at $T_C$ under zero magnetic field around the paramagnetic to ferromagnetic phase transition in both systems. The critical nature and critical components at $T_C$ under out-of-plane magnetic field (*H*//*c*) have been comprehensively investigated [8,9] [10,11].

Fluctuations and correlations in these systems play a decisive role in phase transitions and critical phenomena. In previous studies, the spin-spin correlations above $T_C$ in CGT and CST are clearly revealed in static magnetic susceptibility [7], X-band electron spin resonance [6], second harmonic generation (SHG) [12], thermal expansion [13] etc.. According to the Callen's model, correlation function of spins around $T_C$ can be reflected in magnetoelastic properties due to exchange interactions related spin-lattice coupling [14]. By measuring the thermal expansion $\lambda(T)=dL/L$ with capacitance dilatometry in CGT, the non-zero $d\lambda/dT$ far above $T_C$ due to short range spin-spin correlation was confirmed [13]. Therefore, more elastic properties are expected to show influence by spin-spin correlations in these systems.

In principle, the piezomagnetic coefficient ($=d\lambda/dH$) is also closely related with magnetic interactions and often discussed in terms of strain-dependence of free energy. Clear differences in the magnitude of spin-spin correlation between CST and CST were revealed in ferromagnetic resonance (FMR) measurements [15]. Their critical behaviors around $T_c$ seem different as well. Therefore, the investigation of magnetostrictive properties of CST and CGT may unveil the physics behind their dynamic behaviors, particularly around $T_C$. However, the direct measurement of magnetostriction by capacitance dilatometry is challenging for thin flake

2D materials.

Recently, an ac technique of piezoelectric transducer[16], based on the traditional ME composite configuration [17], is developed to sense the ac piezomagnetic coefficient d$\lambda$/d$H$ of magnetic materials via interfacial strain coupling. This technique has been applied to obtain magnetic phase diagrams of bulk MnSi samples and successfully identified the skyrmion phases and regions with short-range spin-spin correlations [16]. In this letter, we introduce this ME composite technique to study the magnetization dynamics, phase diagrams, and spin-spin correlations of $Cr_2X_2Te_6$ (X=Si, Ge) single crystals. Clear distinction between domain wall motion and domain rotation in magnetization process can be revealed by this method in both systems. Moreover, the comprehensive phase diagrams of both compounds for *H* along *c*-axis and in the *ab*-plane are obtained. Three phase boundaries among ferromagnetic (FM), paramagnetic (PM), and spin-polarized (SP) states, characterized by clear peaks in the real part of ac ME signals, meet at a triple point around $T_c$. The ME signal becomes the maximum at the triple point, implying a significant enhancement of spin-spin correlation at this point. Besides, a stronger spin fluctuation in CST than that of CGT is found in this study.

Single crystals of CGT and CST with a typical size of 9.0×4.5×0.5 $mm^3$ were grown by self-flux method [18]. The single crystal X-ray diffraction is performed (Fig. 1(b)) and the calculated *c*-axis lattice constants of 2.037 nm and 2.032 nm for CGT and CST, respectively, agrees well with the published values. The magnetization is measured in Magnetic Property Measurement System (MPMS, Quantum Design). Magnetostriction is detected by a homemade capacitive dilatometer using a high accurate capacitance bridge (AH2550A, ANDEEN-HAGERLING, INC.). The temperature and magnetic field environment for electrical measurements are supplied by 9 T Dynacool system (Quantum Design). CGT and CST, as the magnetostrictive phases, are glued onto piezoelectric layer of $0.7Pb(Mg_{1/3}Nb_{2/3})O_3$-$0.3PbTiO_3$ (PMN-PT) [001]-cut single crystal, as shown in Fig. 1(c). The magnetoelectric signal of ME laminates $V_{ME}=V_x + iV_y$ where $V_x$ and $V_y$ represent in phase signal and out-of-phase signals respectively, are measured by a homemade setup, same as that in Ref.[16]

Temperature dependent magnetization (M) after field-cooling (FC) of CGT and CST, for magnetic field along *ab*-plane ($H_1$) and *c*-axis ($H_3$) with $H$=0.1 kOe, is presented in Figs. 1(d) and 1(e), respectively. A typical PM to FM phase transition occurs at $T_C$ = 34 K and 67 K for CST and CGT, respectively, consistent with previous reports [11,19]. In $H_3$ configuration, the $H/M$ curves in both compounds deviate from the linear fit below 150 K, confirming a short-range spin-spin correlation in the PM phase far above $T_C$ [7,8]. Moreover, the isothermal magnetization curves are measured for CGT and CST at 30 K and 25 K, respectively [Figs. 1(f) and 1(g)]. The saturation fields of both compounds in $H_3$ configuration (2.0 kOe and 2.5 kOe, respectively) are much smaller than that in $H_1$ configuration (8.0 kOe and 4.5 kOe, respectively), showing a $c$-axis magnetic anisotropy and different magnetization processes [11,19].

We first reveal that the different microscopic domain dynamics in the magnetization processes between $H_1$ and $H_3$ configurations can be reflected in the magnetostriction curve. As an example, the magnetic field and temperature dependence of magnetostriction $\lambda_3$ of CST single crystal for both magnetic field directions are measured, as shown in Figs. 2(a) and 2(c). $\lambda_3$ of CST in both configurations show a clear kink around saturation field, connecting a linear behavior with positive slope for higher field. At lower fields, $\lambda_3$ remains almost zero for $H_3$ configuration but shows a negative quadratic behaviour for $H_1$ configuration. The distinct $\lambda_3(H)$ curves reflect different domain dynamics between two configurations before saturation. In $H_1$ case, Selter *et al*. [7] suggests that the domain rotation process is dominating while for $H_3$, we will show later that it is dominated by domain wall movement, as shown in the insets of Figs. 2(a) and 2(c), respectively. After the saturation, the magnetostriction in the SP phase should be dominated by forced volume magnetostriction governed by exchange energy [20]. From the direct magnetostriction measurements, it is hard to determine the microscopic mechanisms in each configuration. In contrast, the composite magnetoelectric method is a very sensitive ac technique for probing magnetostrictive properties, and thus provides more information on the dynamics of magnetization process.

To apply the composite ME method in both systems, we mechanically bond the single crystals with PMN-PT to form a ME composite, as shown in Fig. 1(c). The PMN-PT with large piezoelectric coefficients is used to enhance the ac magnetoelectric signal $V_{\mathrm{ME}}$ signals. In principle, when dc and ac-driven magnetic fields are applied to the laminate, the $V_{\mathrm{ME}}$ is directly proportional to the piezomagnetic coefficient d$\lambda$/d$H$ of the CGT and CST. In $H_1$ and $H_3$ configurations, we have the following relations [21]:

$$V_{ME33} \propto d_{31} q_{13} \quad (1)$$

$$V_{ME31} \propto -d_{31}(q_{11} + q_{21}) \quad (2)$$

where $V_{\mathrm{ME}ij}$ is the voltage measured along $i$-axis for $H$ applied along $j$-axis, $d_{31}$ is the transverse piezoelectric coefficient, $q_{ij}$ are the piezomagnetic coefficients, i.e., d$\lambda_i$/d$H_j$, where $\lambda_i$ is the magnetostriction along $i$-axis for $H$ along $j$-axis. According to these equations, $V_{\mathrm{ME33}}(H)$ and $V_{ME31}(H)$ should be approximatively proportional to $q_{13}(H)$ and $-[q_{11}(H)+q_{21}(H)]$, respectively.

$V_{ME}$ as a function of magnetic field $H_1$ and $H_3$ for CST at 30 K is measured to compare with $\lambda_3(H)$ curves, as shown in Figs. 2(b) and 2(d). The real part $V_x$ in $H_3$ and $H_1$ configurations shows a peak feature and sign-reversal feature, respectively. The fields where the positive peak appears are in line with the saturation fields in the *M-H* curves. In $H_1$ case, $V_x$ shows a similar profile with that of $q_{31}$. We note that $\lambda_1(H)+\lambda_2(H)+\lambda_3(H)\approx 0$ due to the volume conservation of the sample under the application of the in-plane $H_1$. Then, Eq. (2) can be transformed to $V_{ME31}\propto q_{31}$. In $H_3$ case, $V_x$ shows distinct behaviour with that of $q_{33}$ in the low field region. This may be related to the uniaxial ferromagnetic nature of CST that the domain wall motion process under $H_3$ leads to a negligible $\lambda_3$ and a field dependent $\lambda_1$. Furthermore, the imaginary part $V_y$ can provide extra evidence for the domain wall movement in small $H_3$ (<1.5 kOe). It is well known that the domain wall movement will dissipate energy from depinning behaviors. This leads to a non-zero $V_y$ in the multi-domain state in Fig. 2(b). In contrast, in $H_1$ case, a nearly zero $V_y$ in the measured field region points to the expected non-dissipative domain rotation mechanism. Therefore, the ME signals of the composite are sensitive to the in-plane piezomagnetic coefficients and are suitable to study the CGT and CST single crystals with quasi-2D nature.

The $V_{ME}$ as a function of $H_1$ and $H_3$ for both CGT and CST at other temperatures are shown in supplementary material (see Fig. S1 in supplementary [26]). Below $T_C$, $V_x$ in $H_3$ configuration always shows a peak feature which increases in magnitude and moves to lower fields with increasing temperature. All the peak fields coincide well with the saturation fields obtained from the *M-H* curves. In the imaginary part $V_y$, there are clear peak features in CGT up to 65 K while only weak peaks near $T_C$ in CST, confirming the domain wall movement processes in both compounds in *H*//*c* configuration. Below $T_C$, all $V_x$ in $H_1$ configuration show a sign-reversal feature. The negative peak is weakened while the positive peak is enhanced, and both moves to lower fields with increasing temperature. All the positive peak fields coincide well with the saturation fields obtained from the *M-H* curves. In all the measured temperatures and magnetic fields, negligible $V_y$ for *H*//*ab* (not shown) point to the domain rotation mechanism. Above $T_C$ in the PM phase where no long-range magnetic ordering exists, all the $V_x$ signals still show a large intensity with broad bump features moving to higher fields with increasing temperature. It is counterintuitive to show large magnetostrictive responses in the PM phase and strongly indicates the existence of short-range spin-spin correlation above $T_C$ in both compounds and both configurations.

To further reveal the spin-spin correlations above $T_C$ in these systems, the $V_{ME}$ as a function of temperature under various magnetic fields for $H_1$ and $H_3$ are shown in Fig. 3. All $V_x$ show a pronounced broad peak around $T_C$ which shifts to lower temperatures in small magnetic fields ($H_3$<1 kOe and $H_1$<2 kOe for CGT, $H_3$<2 kOe and $H_1$<1 kOe for CST) and moves towards higher temperatures in stronger fields. This strong peak feature is consistent with the broad bump in magnetic field scan above $T_C$ in both compounds. The non-zero $V_x$ can persist up to two times of $T_C$ in both systems, implying the strong spin-spin correlation up to those temperatures. Besides, we also found shoulder or sign reversal features in $V_x$ below $T_C$ which correspond to the saturation of $M$ in $H$ scan data. The dissipation peaks by domain wall movement are revealed in $V_y$ in $T$ scan data as well. All those features are fully consistent between $H$ scan and $T$ scan data.

Accordingly, the H-T phase diagrams and the microscopic origin of magnetization processes of CGT and CST are constructed from the above measurements, as shown in Fig. 4. Three phases are identified, including paramagnetic phase, ferromagnetic phase with domains and domain walls (DW), and spin polarized (SP) single domain phase. The three boundaries among three phases are determined from the peak position of $V_x$ peaks and are in good agreement with magnetization and ferromagnetic resonance (FMR) measurements [15]. In the $H_3$ configuration [Figs. 4(a) and 4(b)], all three lines converge at one point which is very likely to be a triple point (TP). In the $H_1$ configuration [Figs. 4(c) and 4(d)], there is an extra line within the FM phase which is defined by the negative dips in $V_x$ signals for both $T$ scan and $H$ scan measurements. This line was attributed to a phase boundary between FM and other phases in literature [7]. However, our FMR and magnetization data in $H$ scan support a boundary defined by positive peak in $V_x$ with larger field values [20]. As a result, a TP also exists in $H_1$ configuration for both samples where three boundaries meet at one point. However, the nature of the TP is elusive.

Judging from thermodynamic criteria, the phase boundary between FM and PM is a second order in nature [10,11]; the boundary between FM and SP, in uniaxial magnets for $H$ along easy axis, is also a second order line; but, the boundary between SP and PM phases seems to be a crossover according to the previous specific heat and temperature dependent magnetization measurements in literatures [7]. Also, no enhancement in specific heat around TP is found in CGT [7]. In this sense, the TP is different from the critical point at $H$=0 and $T$=$T_C$ proposed by [10] or a tricritical point at which the order of the transition changes its character from first to second order. In our case, the TP should be a state in which FM, PM, and SP phases coexist at the same time. The physical properties must be exotic at this point like the triple point in water.

To examine the influence of TP on the magnetostrictive properties, we draw the contour plot of $V_x$ –$T$ as the background of each phase diagram. From TP, the $V_x$ becomes smaller towards lower temperatures in the FM phase due to the freezing of spin fluctuation. Towards higher temperatures in the PM phase, $V_x$ slowly decays due to the suppression of short-range spin-spin correlations by thermal fluctuations. With increasing magnetic field from TP, $V_x$ is governed by volume magnetostriction and decays towards higher $H$ due to the increase of Zeeman energy. All the tendencies point to an enhanced $V_x$ signals around the TPs in both systems. The enhancement of ME signal at TP can be understood from its phase coexistence nature. This is very similar to the morphotropic phase boundary in ferroelectrics [22] where the coexistence of competing phases is very susceptible to external stimulus so that an enhanced piezoelectricity is induced. As a result, the largest $V_x \propto d\lambda/dH$ should appear at TP. The enhanced $V_x$ also points to an enhanced spin-spin correlation from the PM phase.

The peak behavior of Vx in T scan data can be understood from the perspective of thermodynamics, where the expression of the piezomagnetic coefficient $\frac{\partial \lambda}{\partial H}$ near $T_C$ is derived by [23]:

$$\frac{\partial \lambda}{\partial H} \approx \frac{\rho M_s}{\rho_0 M_{s0}} \left(\frac{\partial \lambda}{\partial H}\right)_0 + \rho \frac{T}{T_c} \frac{\partial T_c}{\partial p} \frac{\partial M_s}{\partial T} \tag{3}$$

where $\rho$ is density, $M_s$ is spontaneous magnetization, $(\frac{\partial \lambda}{\partial H})_0$, $M_{s0}$ and $\rho_0$ is the value of $\frac{\partial \lambda}{\partial H}$, $M_s$ and $\rho$ at 0 K. Therefore, the divergence of $\frac{\partial M_s}{\partial T}$ around $T_C$ in a second order phase transition from the $T < T_C$ side is responsible for the divergence behaviour in piezomagnetic coefficient and ME signal accordingly [24]. For $T > T_C$ side, spin-spin correlation and the ME signals are related. Based on previous studies, the spin-spin correlation function $<S_f \cdot S_j>$ is proportional to the magnetostriction $\lambda(H, T)$ around magnetic phase transition. As a result, near $T_C$, $\frac{\partial \lambda}{\partial H}$ is governed by spin-spin correlations that [14]:

$$\frac{\partial \lambda}{\partial H} \approx 4 \sum_{j(f,g)} \widetilde{D}_{11}^{\alpha}(f,g) < S_j^z \boldsymbol{S}_f \cdot \boldsymbol{S}_g > \tag{4}$$

where $\widetilde{D}_{11}^{\alpha}(f,g)$ are the phenomenological two-ion magnetoelastic coupling constants between distinct nearest neighbor ion pairs $(f,g)$. $S_j^z$ is the component of spin $j$ along external magnetic field direction and will go over all the spin sites. From Eqs. 1-2, a direct relationship between $V_x$ and $< S_j^z \boldsymbol{S}_f \cdot \boldsymbol{S}_g >$ can be inferred. Note that, for ferromagnets like CGT and CST, $< S_j^z \boldsymbol{S}_f \cdot \boldsymbol{S}_g >$ and $<S_f \cdot S_g>$ is related but different. At zero field, $< S_j^z \boldsymbol{S}_f \cdot \boldsymbol{S}_g >$ is zero in the PM phase since $<S_j^z>$ is zero. While under finite magnetic field, both $< S_j^z \boldsymbol{S}_f \cdot \boldsymbol{S}_g >$ and $<S_f \cdot S_g>$ can persist well above $T_C$ together with different decaying speed [25]. This is due to the combined decay of $<S_j^z>$ and $<S_f \cdot S_g>$ around $T_C$ in $< S_j^z \boldsymbol{S}_f \cdot \boldsymbol{S}_g >$. Finite $<S_f \cdot S_g>$ slowly decays well above $T_C$ in CST and CGT by neutron scattering, SHG, and thermal expansion experiments [8] etc., while $V_x \propto \mathrm{d}\lambda/\mathrm{d}H$ fully decays up to temperatures twice of $T_C$. An enhancement of $< S_j^z \boldsymbol{S}_f \cdot \boldsymbol{S}_g >$ is expected in finite fields around $T_C$. At a fixed temperature in the PM phase, $<S_j^z>$ will increase with the increasing of magnetic field while spin fluctuations will decrease with increasing field. Such contrasting behaviors will lead to a peak feature for $< S_j^z \boldsymbol{S}_f \cdot \boldsymbol{S}_g >$ at finite field around $T_C$ and should be located at TP in CGT and CST.

Finally, our previous studies based on magnetic resonance technique show that the spin fluctuations in CST are much stronger than those for CGT [15]. We will show that the composite ME technique can also compare such fluctuation behaviors around $T_C$ in two systems. The $V_x$ -$T$ data under $H_3$ configuration in both compounds are analyzed by fitting the $V_x$ signals with Lorentz peak functions across the SP-PM and FM-PM phase boundaries, as exemplified in Fig. 4(a) and (b) for CGT under $H_3$=2.2 and 1.5 kOe, respectively. The fitted half width $\Delta T$ of these peaks mark the strong spin-fluctuation regions where the piezomagnetic effect $\mathrm{d}\lambda/\mathrm{d}H$ can still exist away from the two phase-boundaries. Around the TP point, the half width does not show clear anomaly, indicating that the enhancement of spin-spin correlation is localized near this point. To compare the spin fluctuation between two systems, the normalized half widths $\Delta T/T_{\mathrm{peak}}$ of the main peak are drawn as a function of $H_3$ (see Fig. S2 in supplementary [26]). With the field getting stronger, the normalized half width increases consequently, indicating that the spin-spin correlations are constructed in a wider temperature region with the help of magnetic field. Furthermore, the normalized half widths in CST are always larger than that of CGT in the whole field range down to zero field. This is consistent with the expected stronger fluctuation in CST from magnetic resonance measurements [20].

In conclusion, our results demonstrate that the composite ME technique is a highly sensitive probe for domain dynamics and spin-spin correlation and their associated magnetostrictive properties. It is complementary to the capacitance dilatometry method and other methods due to the existence of imaginary part that provides extra information. This technique is particularly superior for studying the quasi-two-dimensional magnetic materials with a thin thickness, because they have a tiny absolute change along thickness direction and is hard to be loaded in the capacitance dilatometry for in-plane measurements. It is also a fast and low-cost technique for mapping the magnetic phase diagram of bulk magnetic materials. We anticipate that this approach can be further applied to investigate spin-spin correlation in other magnetic systems such as antiferromagnets or spin glasses.

## ACKNOWLEDGMENTS

This work is supported by the Natural Science Foundation of China under grant Nos. 11974065, 51725104, 51831003, and National Key R&D Program of China (No. 2021YFB3501402). Y. S. Chai acknowledges the support from Beijing National Laboratory for Condensed Matter Physics. We would like to thank Miss G. W. Wang and Y. Liu at Analytical and Testing Center of Chongqing University for their assistances. We thank X. F. Zhang and Y. Liang for helpful discussion.

## FIGURES

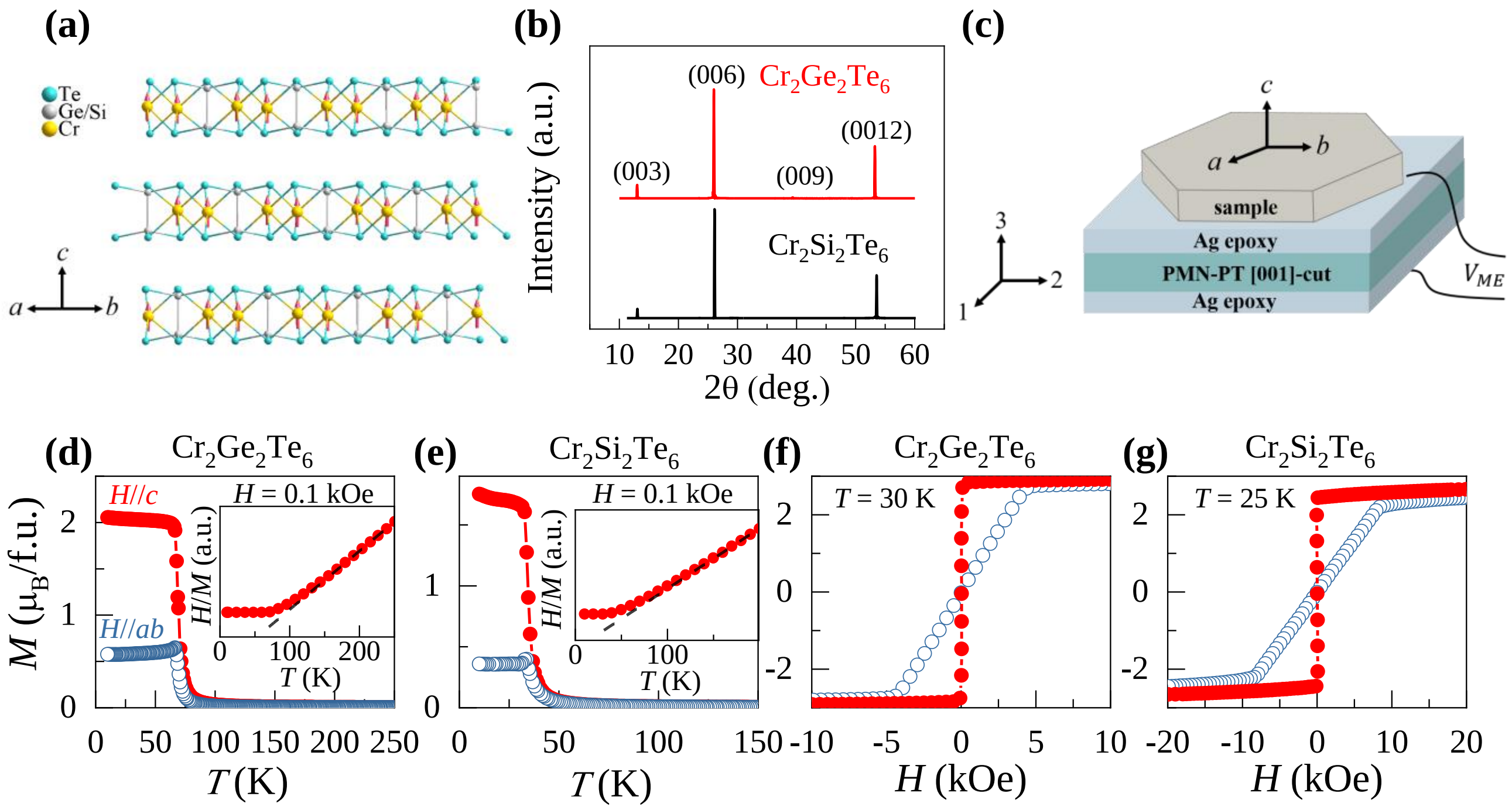

**FIG. 1.** (a) Crystal structure and (b) single crystal X-ray diffraction of $Cr_2Ge_2Te_6$ and $Cr_2Si_2Te_6$. (c) Schematic illustration of the composite ME technique. *T* dependent magnetization of (d) $Cr_2Ge_2Te_6$ and (e) $Cr_2Si_2Te_6$ measured at *H* = 0.1 kOe, and *H* dependence of magnetization of (f) $Cr_2Ge_2Te_6$ and (g) $Cr_2Si_2Te_6$ measured at *T* = 30 K and 25 K respectively, for both *H*//*ab* plane and *H*//*c* axis.

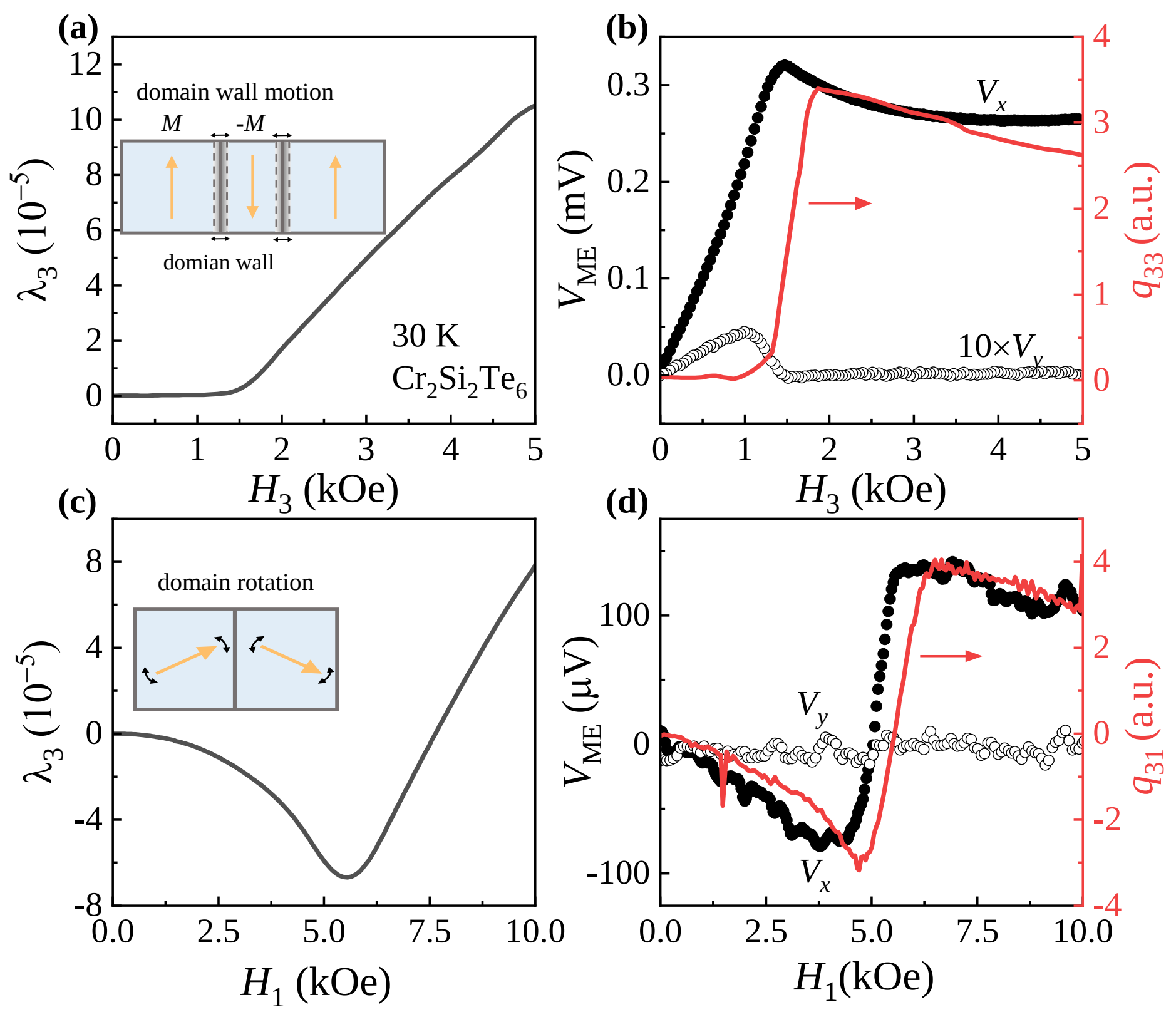

**FIG. 2.** (a) and (c) $H$ dependent magnetostriction $\lambda_3$ for $H//c$ and $H//ab$-plane of $Cr_2Si_2Te_6$ at 30 K. (b) and (d) $V_{ME}$ signal for $H_3$ and $H_1$ configurations at 30 K. The piezomagnetic coefficient $q_{ij}$ is also plotted for comparison.

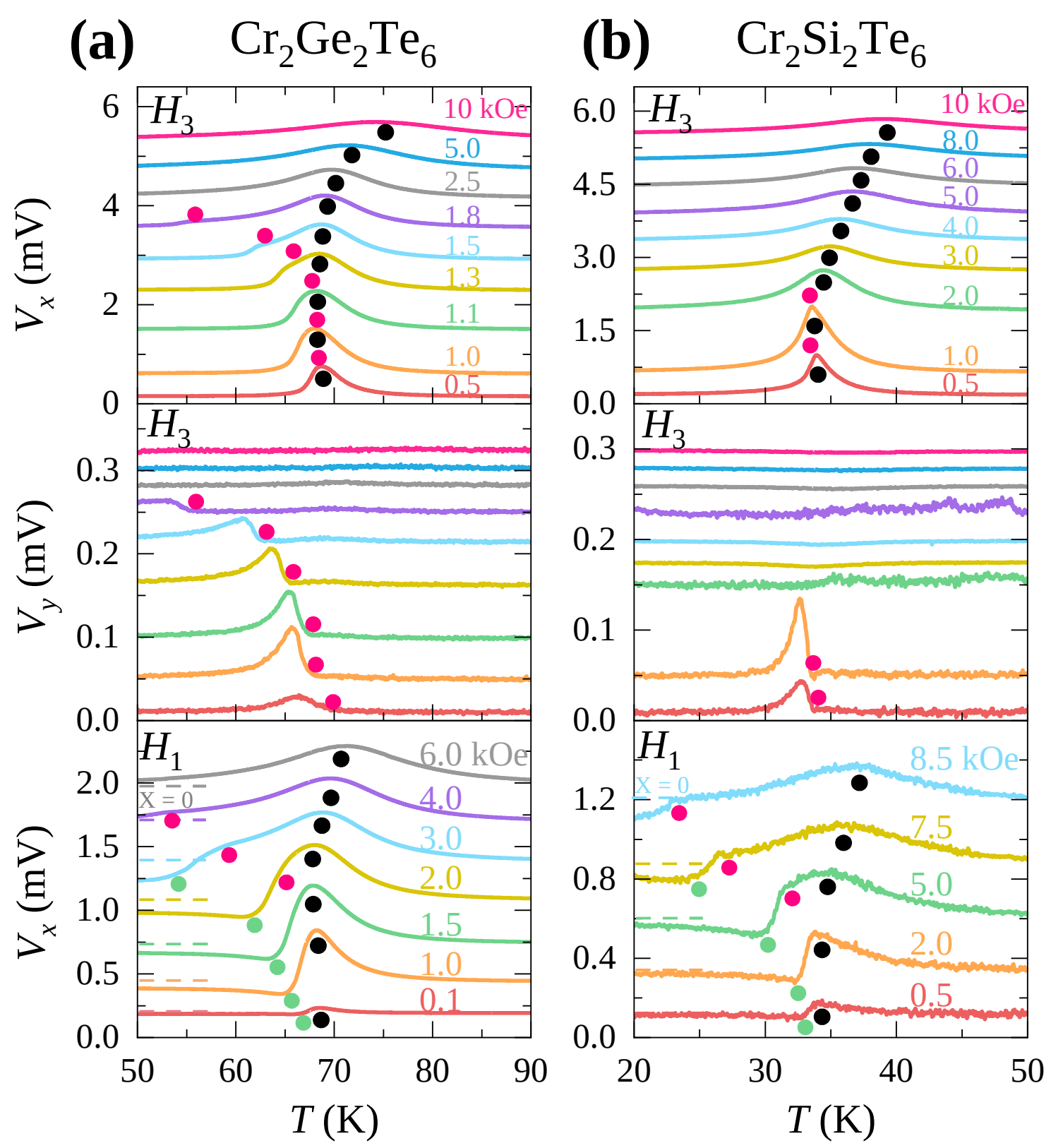

**FIG. 3.** *T* dependent $V_{\mathrm{ME}}$ signals of (a) $Cr_2Ge_2Te_6$ and (b) $Cr_2Si_2Te_6$ measured at selected *H*. The upper and middle panels are the real component $V_x$ and imaginary component $V_y$ in $H_3$ configuration. The lower panels are $V_x$ in $H_1$ configuration. All plots have been shifted vertically for clarity.

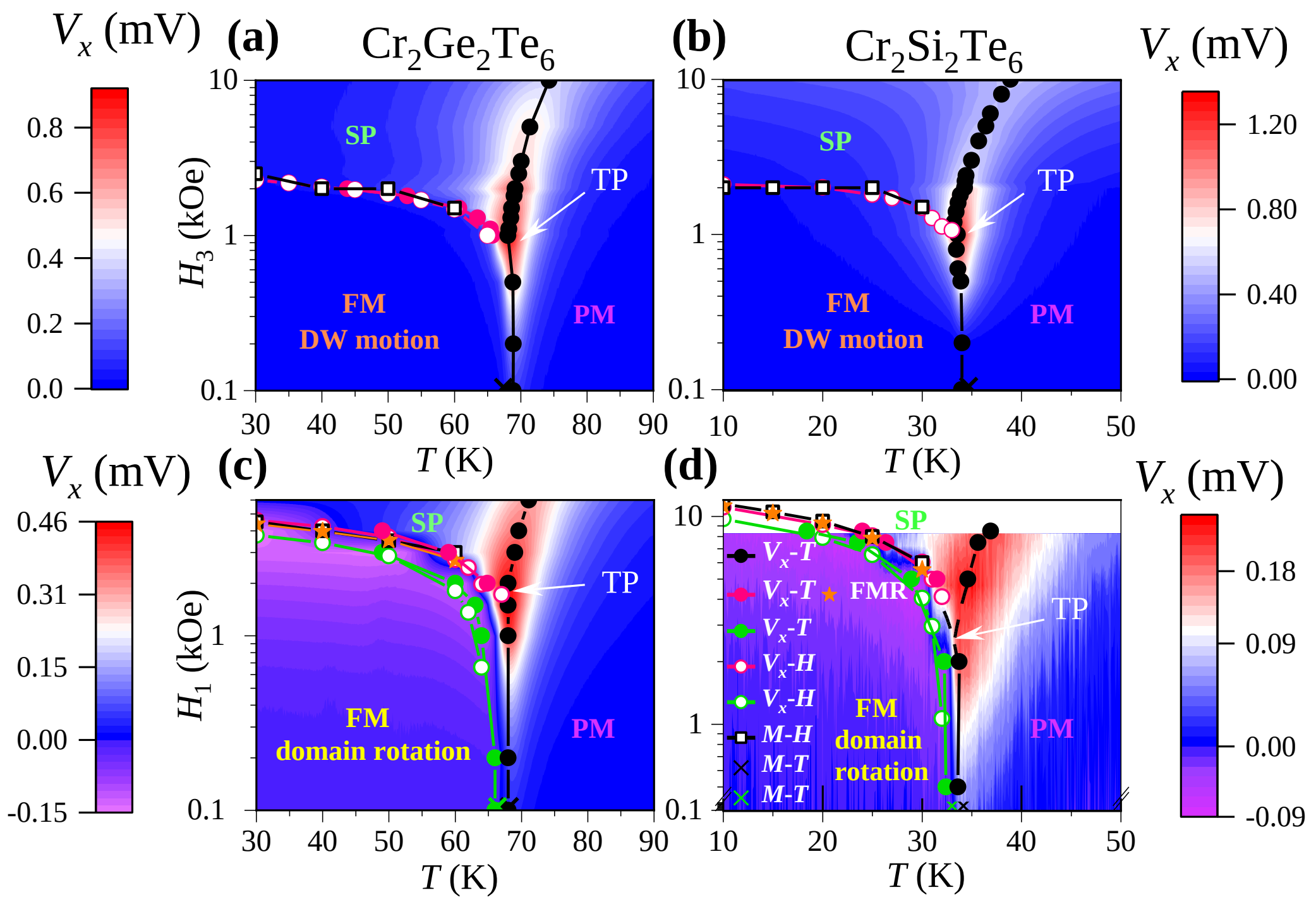

**FIG. 4.** Phase diagram of $Cr_2Ge_2Te_6$ for (a) *H*//*c* axis and (c) *H*//*ab* plane, and of $Cr_2Si_2Te_6$ for (b) *H*//*c* axis and (d) *H*//*ab*-plane. The background of phase diagram is contour plot of $V_x$-$T$. TP represents the triple point and DW means domain wall. The phase boundaries are determined by a variety of measurements.

**Supplementary informations for:**

# Investigation of domain dynamics and spin-correlations in two-dimensional ferromagnetic $Cr_2X_2Te_6$ (X=Si, Ge) by composite magnetoelectric method

Yugang Zhang[1,2†], Zefang Li[3,4†], Jing Zhang[1,2], Long Zhang[1,2], Yonghong Li[1,2], Shuang Liu[2], Young Sun[2*], Wenhong Wang[3,4*], and Yisheng Chai[1,2*]

In this Supporting Information, we provide (1) Field dependence of ME signal for $Cr_2Ge_2Te_6$ (CGT)and $Cr_2Si_2Te_6$ (CST); (2) Field dependent half width of peak in $V_x$-$T$ for CGT and CST.

### Field dependent $V_{ME}$ of CGT and CST

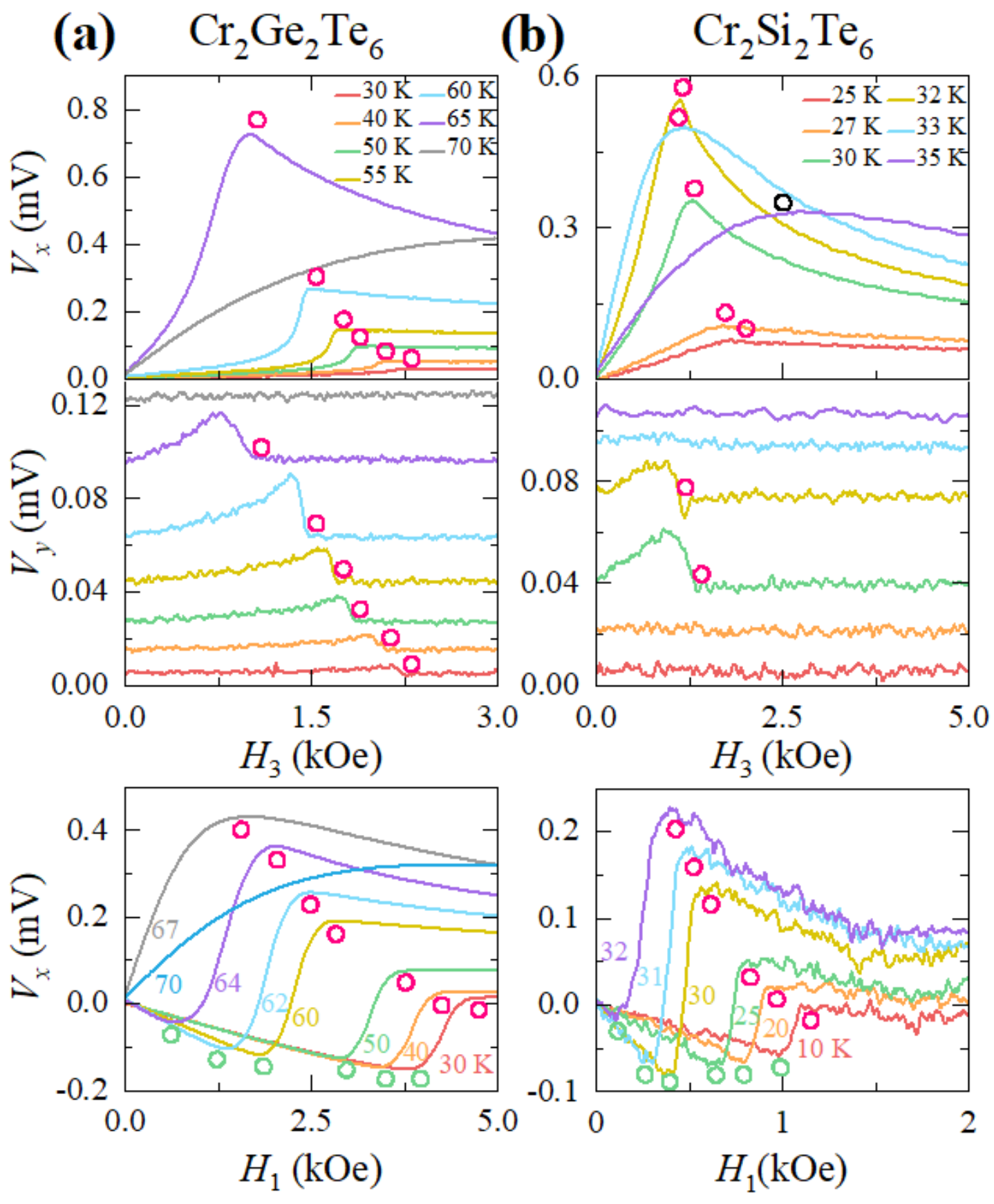

Fig. S1 $H$ dependent $V_{ME}$ signals of $Cr_2Ge_2Te_6$ (a) and $Cr_2Si_2Te_6$ (b) measured at selected $T$, for $H_3$ and $H_1$ configuration. The upper and middle panel of a,b are $V_x$ and $V_y$ in $H_3$ configuration, respectively. The middle panel of a,b has been shifted vertically for clarity.

**H dependent half width of $V_x$-T for CGT and CST**

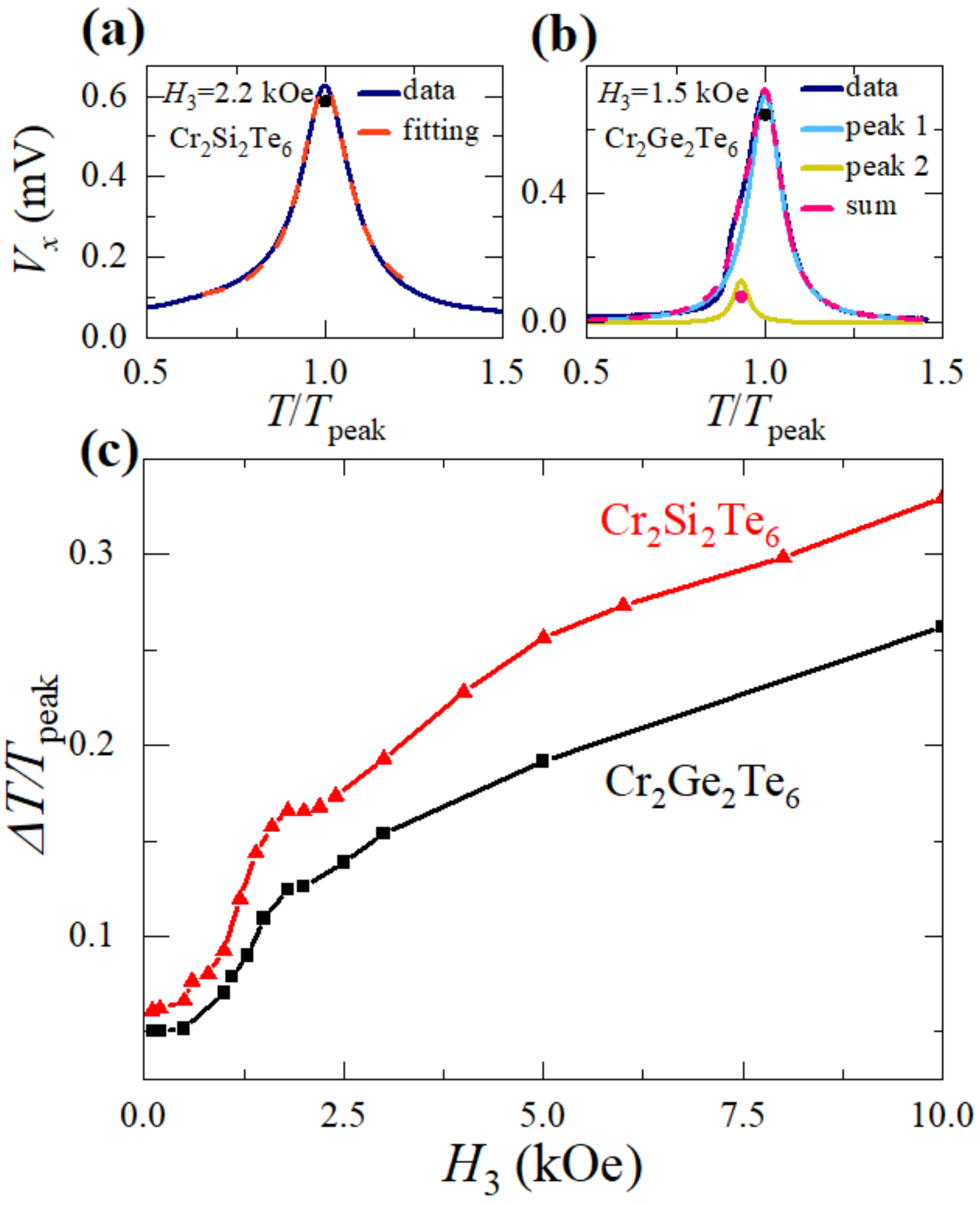

Fig. S2. (c) Magnetic field dependence of half width of $Cr_2Ge_2Te_6$ and $Cr_2Si_2Te_6$, which is extracted from peaks in X-$T$ with magnetic field applied along $c$ axis. (a,b) Examples of Lorentz fitting by which half width is obtained. (a) is the single peak case and (b) is the double peak case.